\documentclass[aps,prl,preprint,superscriptaddress,floatfix]{revtex4-2}
\usepackage{graphicx}
\usepackage{natbib}
\usepackage[english]{babel}
\usepackage{longtable}
\usepackage{xcolor}
\usepackage{hyphenat}
\usepackage{float}

\begin{document}

\title{Dual-adatom diffusion-limited growth model\\ for compound nanowires: Application to InAs nanowires}

\author{Danylo Mosiiets}
\author{Yann Genuist}
\author{Joël Cibert}
\affiliation{Univ. Grenoble Alpes, CNRS, Grenoble INP, Institut Néel, 38000 Grenoble, France}
\author{Edith Bellet-Amalric}
\affiliation{Univ. Grenoble Alpes, CEA, Grenoble INP, IRIG, PHELIQS, 38000 Grenoble, France}
\author{Moïra Hocevar}
\email[]{moira.hocevar@neel.cnrs.fr}
\affiliation{Univ. Grenoble Alpes, CNRS, Grenoble INP, Institut Néel, 38000 Grenoble, France}

\begin{abstract}

We propose a dual-adatom diffusion-limited model for the growth of compound semiconductor nanowires \textit{via} the vapor-liquid-solid or the vapour-solid-solid mechanisms. The growth is catalyzed either by a liquid or a solid nanoparticle. We validate the model using experimental data from the growth of InAs nanowires catalyzed by a gold nanoparticle in a molecular beam epitaxy reactor. Initially, we determine the parameters (diffusion lengths, flux to the seed, Kelvin effect) that describe the growth of nanowires under an excess of one of the two beams (for instance, group III or group V atoms). The diffusion-limited model calculates the growth rate resulting from the current of atoms reaching the seed. Our dual-adatom diffusion-limited model calculates for a compound semiconductor, the instantaneous growth rate resulting from the smallest current of the two types of atoms at a given time. We apply the model to analyze the length-radius dependence of our InAs nanowires for growth conditions covering the transition from the As-limited to the In-limited regime. Finally, the model also describes the complex dependence of the transition between both regimes on the nanowire radius and length. This approach is generic and can be applied to study the growth of any compound semiconductor nanowires.

\end{abstract}

\maketitle

\section{Introduction}
Nanowire growth began in the early 1960s with silicon. Experimental results  \cite{Wagner1964} and theoretical studies \cite{Ruth1964} during this period identified the role of the diffusion of single types of adatoms in the growth process. However, various mechanisms have been and are currently being invoked and discussed based on the impact of the nanowire diameter on the growth rate \cite{Schmidt2007}. During the following decades after, interest in the growth of nanowires made of compound semiconductors has progressively increased, particularly III-V nanowires such as GaAs and InAs \cite{Hiruma1995,Ohlsson2001,Harmand2005}. For instance, GaAs nanowires provide an alternative to planar structures for integrating photonics on Si substrates, and InAs nanowires stand out for their advantages in applications such as chemical sensors \cite{Tseng2017} and photodetectors \cite{Xu2020}. Today, InAs nanowires contribute significantly to quantum devices because of their small bandgap, large mobility and large g-factor. InAs nanowire devices feature not only high mobility transistors \cite{Konar2015} but also serve as a promising platform for various quantum bit implementations, including spin-based quantum bits (qubits) \cite{Nadj2010} and transmon qubits \cite{Larsen2015}.

All those novel devices were enabled not only thanks to better cleanrooms and better lithography techniques, but also thanks to better materials. Knowledge and control over growth and crystalline quality have been foundational to the improvement of nanowires. InAs nanowires grow self-catalyzed and by the gold-assisted vapour-solid-solid (VSS) or vapour-liquid-solid (VLS) mechanisms. In the latter case, they are particularly easy to grow thanks to the large range of parameters for which the VLS mechanism works. Parameters are either external (In and As fluxes sent by the cells, growth temperature, growth time) or specific to the sample (size of the Au nanoparticle, substrate orientation, distance between nanoparticles, surface chemistry). 

Understanding and controlling the growth of nanowires is \textit{a priori} more complex for a compound semiconductor than for an elemental semiconductor since it depends on the behavior of the two species. Many studies focused on growth conditions under a strong excess of one constituent, so that the growth is essentially controlled by the other one. For instance, the growth of GaAs nanowires self-catalyzed by a Ga nanoparticle usually reveals the behavior of the As molecules and adatoms, albeit complex \cite{Glas2013b}. As detailed below, arsenic is highly volatile, and its adatoms exhibit a very short diffusion length, leading to a weak dependence of the growth rate on the nanowire radius. In the opposite case of a strong arsenic excess, the negligible evaporation of the type-III adatoms and their diffusion length in the micrometer range result in a characteristic decrease of the growth rate upon increasing the nanowire radius \cite{Jensen2004}. However, a cutoff at small radius is sometimes reported \cite{Froberg2007} and attributed either to a size effect ("Gibbs-Thomson effect") incorporated into the single-species model, or to the low probability of nucleation on a small area \cite{Dubrovskii2014}. We may note here that the notion of a diffusion length, $\lambda=\sqrt{D \tau}$ may cover different mechanisms: $D$ is a phenomenological diffusion coefficient, and the residence time $\tau$ may correspond to the desorption of the adatoms or to their incorporation. Fewer studies take into account the two species. An experimental diagram of the growth rate of InAs nanowires seeded by Au colloids of radius 10 nm, \textit{vs.} As and In beam equivalent pressures (BEP), was reported in Ref. \cite{Babu2011}, and described qualitatively. Indeed, both In and As BEPs play a significant role. Quantitative models were proposed more recently \cite{Dubrovskii2014b,Johansson2019,Dubrovskii2020}. They confirm that in case of a strong excess of one species, the other one controls the growth. They show also that the dependence of the chemical potentials, and hence the nucleation rates, on both fluxes may make the transition between the two limits quite progressive.

Here, we propose to look deeper into the influence of the flux of each type of atoms, for instance type-III and type-V fluxes, on the growth of compound nanowires, both experimentally on InAs nanowires, and theoretically. InAs is particularly well suited for this study because In and As behave very differently (In doesn’t re-evaporate, has long diffusion length and a small direct flux; As re-evaporates, suffers from the Kelvin effect, has limited diffusion and is re-emitted from surfaces). We synthesized nanowires from gold catalysts of various radii, using different In and As cell BEPs, at a temperature chosen within the range ($\mathrm{400^\circ \text{C}-420^\circ \text{C}}$) where the growth rate of InAs nanowires was shown to reach a maximum \cite{Tchernycheva2007}. We found that the nanowire growth rate shows a maximum for a given radius for all values of the In and As BEPs. Our scan of the As BEP allows us to observe the evolution of the length dependence on radius from a dependence typical of an As-limited growth, to a dependence typical of the growth limited by the In flux. However in all cases no nanowires are observed below a critical radius, the value of which depends slightly on the In flux and not on the As flux. We then propose a dual-adatom diffusion-limited growth model that takes into account the dynamics of both type-III and type-V adatoms, based on their very different properties, and on a size effect which is discussed as the Kelvin effect acting on As only. Our model shows that for a given V/III ratio, the growth rate can be limited by either one of the two adatom currents, depending on the diameter and the length of the nanowire. As the currents of both type-III and type-V adatoms are not balanced, we can describe solely their limiting behaviors. This enables the calculation for the overall growth of our nanowires by applying these limiting behaviors at each moment of growth. Finally, we find a set of parameters that allows us to fit all our experimental data. The diffusion lengths of In and As on the nanowire sidewall facets are approximately 7000 nm and  100 nm, respectively, at $\mathrm{420^\circ \text{C}}$.

The paper is organized as follows: first, we present the growth conditions of the nanowires and the results of the experimental growth of InAs nanowires. Then, we present our nanowire growth model with a particular focus on III-V compound semiconductors. We demonstrate the dependence of the instantaneous growth rate of each species on the nanowire length and apply the model to our experimental data with nanowires grown under various conditions. Furthermore, we discuss our experimental results that cover the transition between As-limited and In-limited growth, our dual-adatom diffusion-limited model, and finally, the effect of a critical radius characterizing the sharp decrease in growth rate at a small radius.

\section{Experimental results}

\subsection{Growth conditions and characterization}

 InAs nanowires are synthesized utilizing the VLS mechanism, assisted by gold catalysts, within a solid-source molecular beam epitaxy (MBE) reactor. Sample preparation proceeds as follows: several pieces of InAs(111)B substrates are de-oxidized in a solution of $\mathrm{H_3NO_4:H_2O}$ (1:1) for 5 to 10 minutes, followed by a thorough rinsing in de-ionized water for an additional 10 minutes. Subsequently, Au colloids (purchased from BBI Solution) with radii ranging from 10 to 50 nm (with a size dispersion of 2\%) are deposited onto the InAs substrates via drop-casting. For each growth experiment, a minimum of four colloid nominal sizes are assessed.

The samples are then glued on a Si substrate and mounted onto a molybdenum block, which is subsequently introduced into the MBE reactor. An initial annealing step is performed at $\mathrm{250^\circ \text{C}}$ under a vacuum of $\mathrm{10^{-9}}$ Torr, to eliminate condensed water. Furthermore, the molybdenum block is introduced to the main chamber for growth. The substrate temperature is raised to $\mathrm{500^\circ \text{C}}$ to form the Au-In eutectic under an As BEP of $\mathrm{10^{-5}}$ Torr measured by a Bayard-Alpert gauge. The As cracker cell is operated at reservoir and cracker temperatures of $\mathrm{350^\circ \text{C}}$ and $\mathrm{650^\circ \text{C}}$, respectively  so that the beam is mainly composed of As$_4$ molecules \cite{Campion2010}. 

After the annealing step, which lasts around 10 min, the sample is cooled down to the growth temperature ($\mathrm{420^\circ \text{C}}$). At this temperature, after 5 min of temperature stabilization during which the As BEP is set to the desired value and the In shutter is opened in order to prevent a burst, growth is initiated by opening the main shutter. Growth is conducted at constant In and As BEPs. After 30 min of growth, the main shutter is closed and the sample returned to room temperature.

\begin{table}

\caption{Samples for the study of the As BEP dependence.\label{table1}}
\begin{ruledtabular}
\begin{tabular}{lcccc}
%	&NW898&	NW899&	NW900&	NW901\\
	&I25R60&	I25R20&	I25R10&	I25R05\\ \hline
In BEP  (x$\mathrm{10^{-7}}$ Torr)&2.5&	2.5&	2.5&2.5\\
As BEP  (x$\mathrm{10^{-6}}$ Torr)&15&	5& 2.5	&1.25\\
As/In BEP ratio&60&	20&	10	&5\\
&&&&	\\
Projected In flux (nm/s)&0.16&	0.16&	0.16	&0.16\\
Projected As flux (nm/s)&1.65&	0.55&	0.28	&0.14\\
Projected As/In flux ratio&11&	4&	2	&1\\
&&&&	\\
Growth time (min)&30&	30&	30	&30\\	 	
\end{tabular}
\end{ruledtabular}

\end{table}

Once removed from the MBE reactor, the samples are imaged in a scanning electron microscope (SEM) at a 45$^\circ$ tilt. We use a Zeiss Ultra$+$ Field Emission SEM with an acceleration voltage of 5 keV. Broad views of the samples and zoomed views of single nanowires are recorded. The diameters of the nanowires are measured at the basis and at the tip of the nanowire, just below the catalyst. This is done on images with a magnification high enough to obtain an accurate value of the diameter. The length of the nanowire is measured from the basis (at the top of the pyramid which usually forms around it), to the top of the nanowire (below the catalyst-nanowire interface), on images with a lower magnification. We measure a minimum of 30 nanowires per colloid size to ensure a reliable statistic of the sample. The nanowire height is corrected by adding the mean thickness of the two dimensional layers calculated from the growth rate calibrations. We consider this thickness to be a sufficient approximation for the pyramid's height.

The values of the flux emitted by the cells have been calibrated through the measure of the growth rates of InAs layers on a (001) InAs substrate, using RHEED oscillations. In the following, the term "flux" will be used for the flux of atoms, projected onto the substrate surface, and expressed as the growth rate of InAs in nm/s. The values are given in Tables~\ref{table1} and ~\ref{table4}. Note that we probably underestimate slightly the As flux. For more details, see Supplementary Material S1 in ref~\cite{suppldata2024}. As the angle of incidence with the substrate is larger for our As beam than for our In beam, for a given value of the As-to-In projected flux ratio, the corresponding ratio for the flux along the cell axis is larger, and the ratio of flux to the nanowire sidewalls is even larger.

\subsection{Results}
\begin{figure}[H]
\centering
\includegraphics[page=1,trim=0cm 0cm 0cm 0cm, clip, width=8cm]{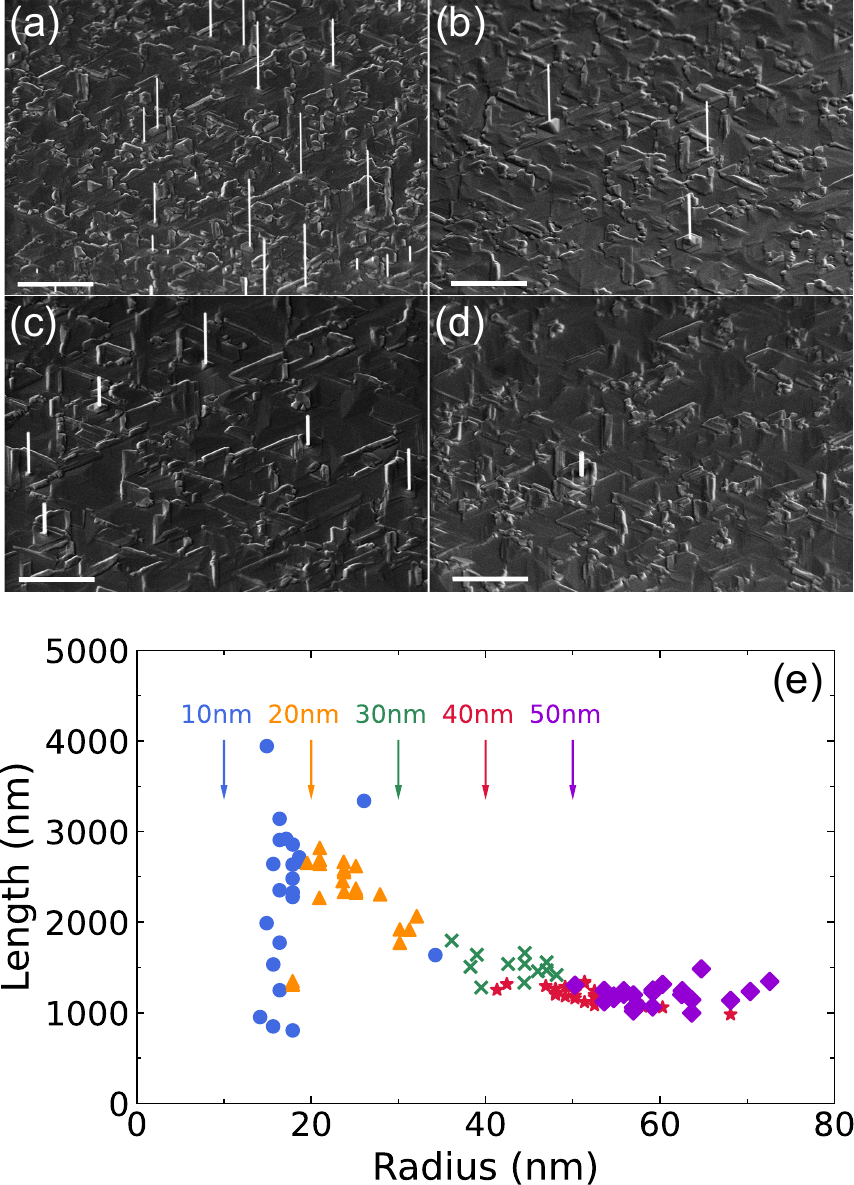}

\caption{Au-assisted InAs wires grown on an InAs (111)B substrate at large As flux (sample I25R60, As BEP=15$\times 10^{-6}$ Torr, In BEP=2.5$\times 10^{-7}$ Torr, As/In flux ratio of 11). (a) to (d), SEM images at 45$^\circ$ tilt. Scale bars are 2 $\mu$m. Radius of colloids: (a) 10~nm, (b) 20~nm, (c) 40~nm, (d) 50~nm. (e) Nanowire length \textit{vs.} nanowire radius. The initial radius of the colloids is displayed with an arrow, the corresponding nanowires are identified by the color of the symbols.\label{fig:1}}
\end{figure}

We show an example of such nanowires grown in our MBE system with a V/III flux ratio of 11 (corresponding to a BEP ratio of 60) using colloids with sizes ranging from 10 to 50 nm as initial radius (Fig.~\ref{fig:1}(a) to~\ref{fig:1}(d)). The nanowires are cylinder-shaped, with the same radius at the bottom and at the tip, equal or slightly smaller that of the gold particle. There is no evidence of radial growth that could result in a pencil-shaped \cite{Tchernycheva2007,Dayeh2009a} or tapered (conical) \cite{Babu2011,Dayeh2009a} nanowire. It is clear from the images that the nanowires grown from 10-nm radius colloids show a large dispersion in length (Fig.~\ref{fig:1}(a)), while the larger radii nanowires are shorter and more homogeneous in length.

The length-radius dependence is shown in Fig.~\ref{fig:1}(e). A salient feature is that each colloid size, symbolized by the colour of the symbols and marked by arrows, gives rise to a range of the nanowire radius much broader than the initial range of the colloid radius. This agrees with previous reports. Smaller radii have been ascribed to splitting \cite{Jung2014} or secondary nucleation of In nanoparticles \cite{Dubrovskii2016b}. Larger radii result from coalescence \cite{Gomes2015}. The effect of coalescence is particularly obvious for the smallest crystallites but we could not observe any trace left behind by the moving nanoparticle, as reported for the growth of GaAs nanowires \cite{Zhang2009}. As a result, we obtain a dependence of the length over a continuum of the radius value, with no clear dependence on the initial crystallite size. It features an extremum at radii less than 20 nm. Above 20 nm, the length of the nanowires decreases with increasing radius. As discussed later on, such a behavior is generally attributed to a growth limited by diffusion with a large diffusion length. As in Ref.~\cite{Froberg2007}, we also observe a sharp cut-off below which the nanowires do not grow.

\begin{figure}
\centering
\includegraphics[page=1,trim=0cm 0cm 0cm 0cm, clip, width=8cm]{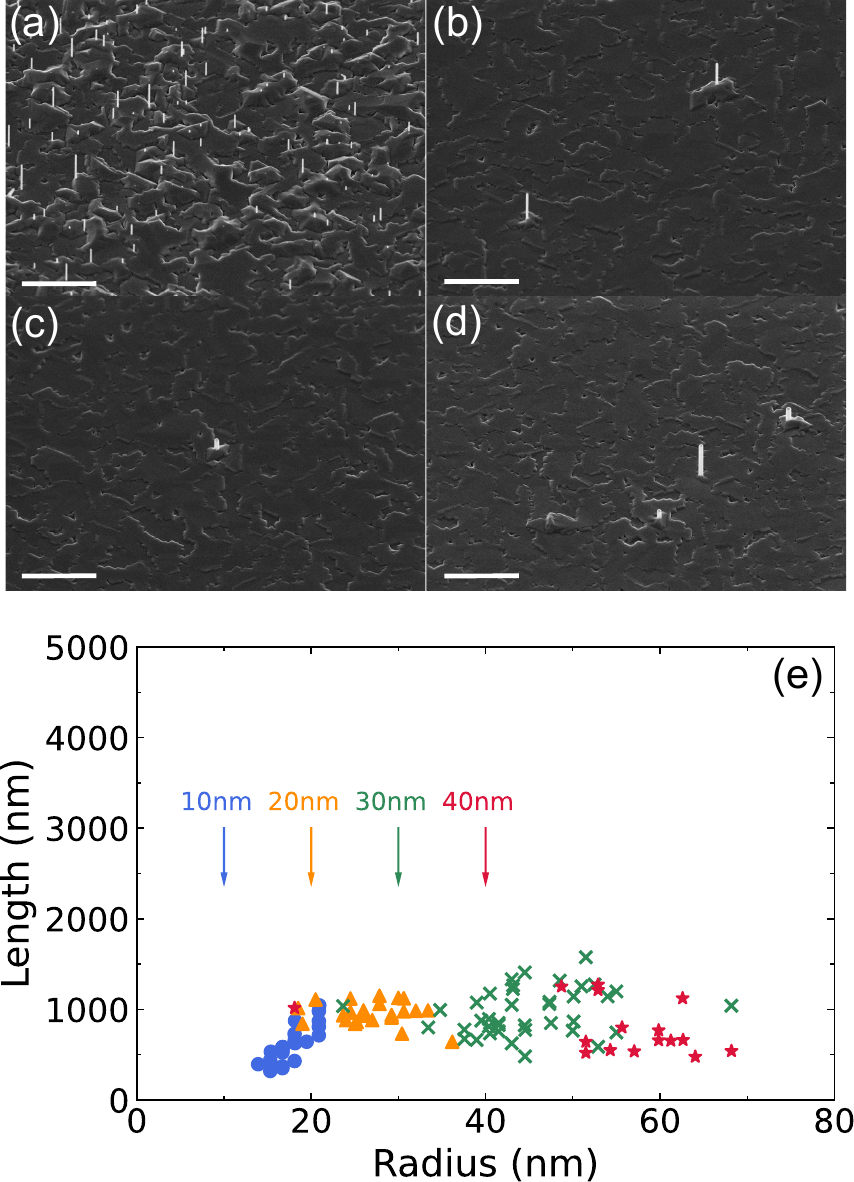}

\caption{Same as Fig.~\ref{fig:1} for a low As flux (sample I25R05, As BEP= 1.25 $\times 10^{-6}$ Torr, In BEP=2.5$\times 10^{-7}$ Torr, As/In flux ratio=1).\label{fig:2}}
\end{figure}

\begin{figure*}

\centering
\includegraphics[page=1,trim=0cm 0cm 0cm 0cm, clip, width=15cm]{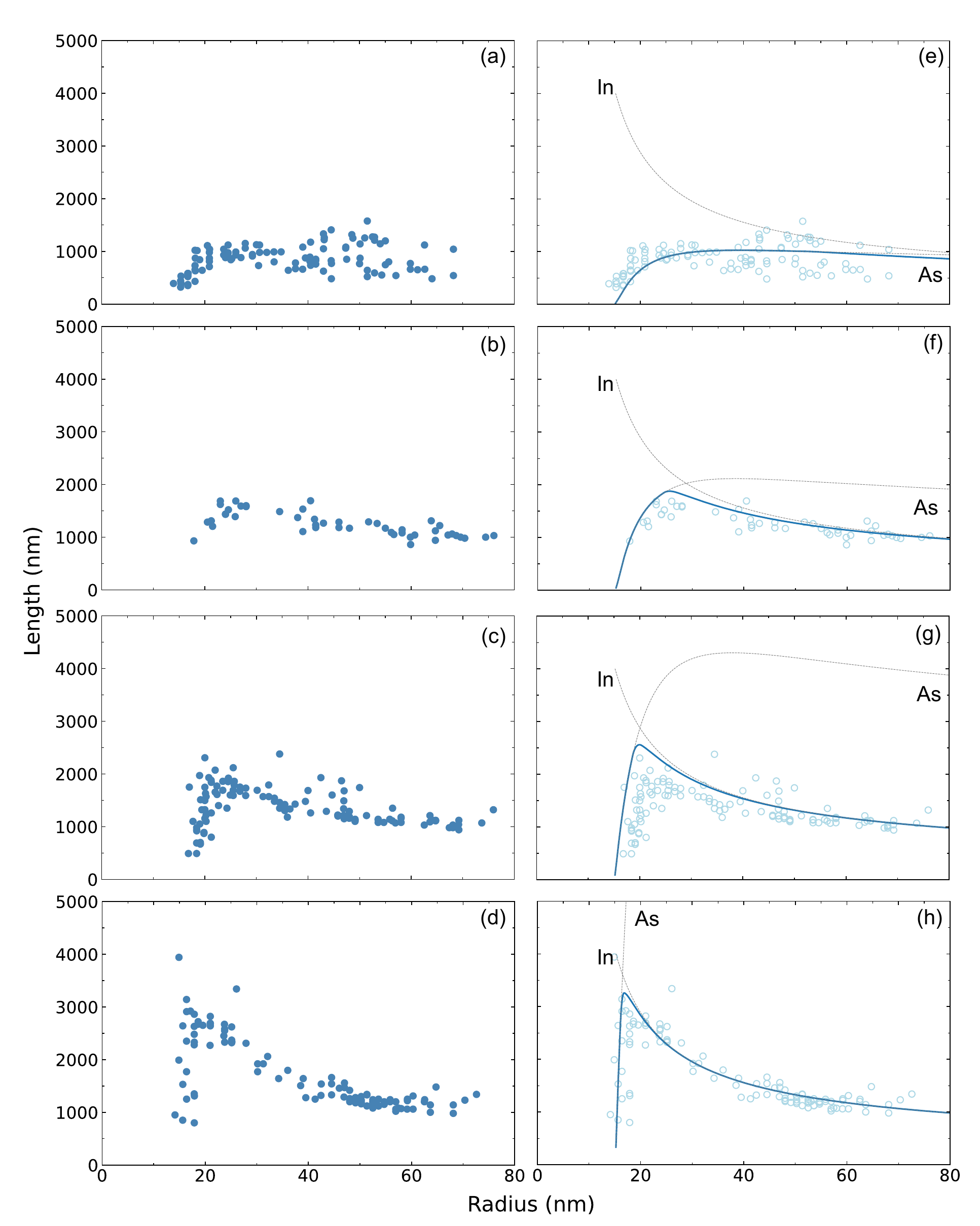}
\caption{Length-radius dependence for different values of the As flux. The symbols show the experimental data at an As/In flux ratio equal to (a) 1, sample I25R05; (b) 2, sample I25R10, same as in Fig.~\ref{fig:2}; (c) 4, sample I25R20 and (d) 11, sample I25R60, same as in Fig.~\ref{fig:1}. (e)-(h) Same as (a) to (d) with the solid blue lines showing the fit calculated using the model. The grey dotted lines are the limits due to In and As, respectively.}\label{fig:3}
   
\end{figure*}

Next, we assess how reducing the As beam equivalent pressure impacts the growth rate of the nanowires. To do this, we grow a series of samples using identical growth conditions to the previous sample, with the exception of lowering the As BEP. We use V/III flux ratios of 4, 2, and 1, respectively. Fig.~\ref{fig:2} illustrates the behavior of a sample grown under a V/III flux ratio of 1 (see the additional samples in  Fig.~S2 ~\cite{suppldata2024}). The nanowire shape is unchanged, see Fig.~\ref{fig:2}(a) to (d). For all the samples of the series, there is an extremum in the dependence of the length on the radius; however, its magnitude increases and shifts towards smaller radii as the V/III flux ratio increases (Fig.~\ref{fig:3}). Also, the increase in nanowire length becomes more and more noticeable as the radius decreases further. The cut-off is still present and at the same radius.

\begin{figure}[H]
\centering
\includegraphics[page=1, trim=0cm 0cm 0cm 0cm, clip, width=8cm]
{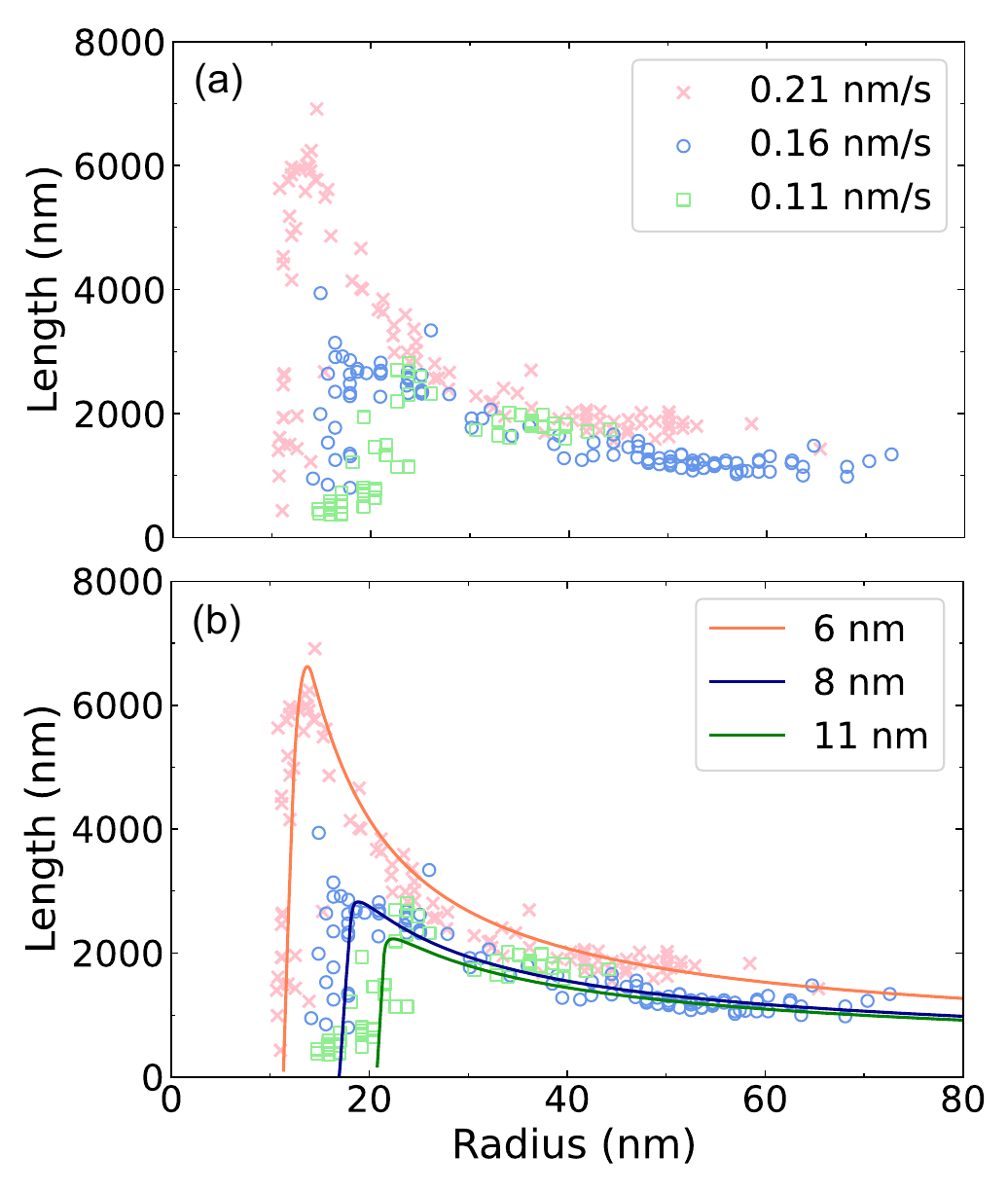}%
\caption{(a) Length-radius dependence for different values of the In flux. The symbols show the experimental data at an In growth rate equal to 0.11 nm/s (sample I18R100), 0.16 nm/s (sample I25R60), and 0.21 nm/s (sample I38R42).  Note the 30\% larger growth time for sample I18R100 (b) Same as (a) with the solid lines calculated using the model and the parameters of Tables~\ref{table3}, \ref{table4}, but for fitting values for $R_0$  (Table~\ref{table6}).} \label{fig:4}
\end{figure}

Nanowires grown with a different In flux (increased or decreased by a factor of 1.5, respectively) still have a cylinder shape (see Fig.~S4 ~\cite{suppldata2024}). Examples of their length \textit{vs.} radius dependence are shown in Fig.~\ref{fig:4}(a). The most noticeable change is a decrease of the cut-off radius upon increasing the In flux.

\begin{table}[H]

\caption{Samples for the study of the In BEP dependence.\label{table4}}
\begin{ruledtabular}
\begin{tabular}{lccc}
	%&NW934&		NW898&	NW939\\
	&		I18R100&I25R60&	I38R42\\ \hline
In BEP (x$\mathrm{10^{-7}}$Torr)&1.8	&2.5&	3.8\\
As BEP (x$\mathrm{10^{-6}}$Torr)&18&15&	16\\
As/In BEP ratio&	100&60	&42\\
&&&	\\
Projected In flux (nm/s)&	0.11&0.16	&0.21\\
Projected As flux (nm/s)&	1.99&1.65	&1.77\\
Projected As/In flux ratio &18&	11	&8\\
&&&	\\
Growth time (min)&	40&30	&30\\	 	 	 
\end{tabular}
\end{ruledtabular}
\end{table}

\section{Growth model}

\subsection{III-V nanowires growth model}
The VLS (and the VSS as well) growth of compound semiconductor nanowires takes place because two types of atoms (such as type-III and type-V) are incorporated into the seed nanoparticle. The nanoparticle is liquid in the case of the VLS growth, while it is solid in the case of the VSS. Hence the relevant quantities are not the flux from the cells, but the currents to the seed, due to diffusion of adatoms from the substrate ($J_{sub}$), diffusion of atoms impinging the sidewall facets ($J_{f}$), direct impingement onto the seed ($J_{Au}$), and re-evaporation ($J_{evap}$):  $J=J_{sub}+J_f+J_{Au}-J_{evap}$. 

In a simple approach, $J$ is determined for each species by solving a diffusion equation for the adatom density on the substrate and along the nanowire sidewall facets, assuming that the seed is a perfect trap \cite{Dubrovskii2006b}. In a one-species model, the growth rate of the nanowire of radius $R$ is$:\frac{dL}{dt}=\frac{J}{\pi R^2}$.

In our model, we evaluate separately the diffusion-limited currents of type-III and of type-V atoms, $J_{III}$ and $J_{V}$, under the fluxes $F_{III}$ and $F_{V}$, and assume that the nanowire instantaneous growth rate is determined by the smallest current: 

\begin{equation}\label{Eq1}
\frac{dL}{dt}=\frac{\min (J_{III},J_{V})}{\pi R^2} .
\end{equation}

It is quite convenient  \cite{Dubrovskii2006b} to consider the growth rate of a two-dimensional layer under the same flux $F$ from the cell, $\frac{dH}{dt}=F \cos \alpha $, where $\alpha$ is the incidence angle of the incoming beam with respect to the normal to the substrate. This results in expressing the nanowire growth rate as $\frac{dL}{dH}$. 
 
\textbf{Type-III adatoms} are known to be non-volatile \cite{Sibirev2012,Koryakin2019a,Koryakin2019b} at the usual growth temperatures. Hence, we let $J_{III_{evap}}=0$. This assumption can be revisited if the growth temperature is high, or for different materials (the evaporation rate of Ga from the seed corresponds to 12\% of the deposition rate at most \cite{Dubrovskii2006a}). When sending a quantity of type-III atoms corresponding to a 2D layer thickness $dH_{III}=F_{III} \cos \alpha_{III} dt$, the maximum increase in the nanowire length, $dL$,  is obtained \cite{Dubrovskii2009a} as follows:

\begin{widetext}
\begin{equation} \label{Eq2}
dL_{III}=\left[\frac{2}{R} \frac{1}{U'(\frac{L}{\lambda_{f,III}})}\lambda_{s,III}  \frac{K_{1}(\frac{R}{\lambda_{s,III}})}{K_{0}(\frac{R}{\lambda_{s,III}})} + \frac{2}{R} \frac{U(\frac{L}{\lambda_{f,III}} )}{U'(\frac{L}{\lambda_{f,III}})} \frac{\tan\alpha_{III}}{\pi}\lambda_{f,III} +(1+\kappa_{III})\right] dH_{III}
\end{equation}
\end{widetext}

The first term in the right-hand side of Equation \ref{Eq2} is due to the flux to the substrate diffusing to the nanowire, the second one to the flux to the nanowire sidewalls, and the third one to the flux to the seed. The flux is characterized by its intensity $F_{III}$ and the incidence angle $\alpha_{III}$ of the incoming beam with respect to the normal to the substrate, and $dH_{III}=F_{III} \cos \alpha_{III} dt$. The diffusion lengths of the type-III adatoms are written $\lambda_{f,III}$ and $\lambda_{s,III}$ for diffusion along the nanowire sidewall facets and on the substrate, respectively. The two dimensional character of the diffusion on the substrate is taken into account by the ratio of Bessel functions of the second kind of order $i$, $K_i(x)$, evaluated at $x=\frac{R}{\lambda_{s,III}}$. The lattice parameters (unit cell volume, surface unit cell area) are incorporated into the diffusion lengths. Finally, the functions $U(x)$ and $U'(x)$ are given by $U(x) = \sinh x + \frac{\lambda_{f}}{\lambda_{s}} \frac{K_1}{K_0}(\cosh x - 1)$ and $U'(x) = \cosh x + \frac{\lambda_{f}}{\lambda_{s}} \frac{K_1}{K_0} \sinh x$, respectively. The direct flux to the seed contains a parameter $\kappa_{III}$ which takes into account the shape of the seed and the incidence of the beam \cite{Glas2010c}. It vanishes when the seed is nearly flat or when it adopts a shape resembling a half-sphere at small angles of $\alpha_{III}$.

\textbf{Type-V atoms} are known to be highly volatile, and to feature a diffusion limited in length, if any. Then Eq. \ref{Eq2} has to be modified. %have a limited diffusion length?
\begin{itemize}
    \item The current into the seed is shared between the growth of the nanowire, and the re-evaporation ($J_{V_{evap}}\neq0$). Thus, we rewrite the flux equation for type-V adatoms as $J=\left(J_{Sub}+J_f+J_{Au}\right) (1-\theta_{evap}^{in})$, where $\theta_{evap}^{in}$ is the ratio of the evaporation current to the input current. Similar terms were derived in Ref.~\cite{Dubrovskii2014b} at every point of the nanowire-seed system, and shown to be related to the chemical potential. As discussed below, we will keep the term $\theta_{evap}^{in}$ fixed when the current from the group V is the minority current, and keep only the factor $\exp(\frac{R_0}{R})$ which describes size effects.
    \item A portion of the flux reaching the substrate is re-emitted, leading to a significant flux towards both the nanowire sidewalls and the seed \cite{Pishchagin2021}. This can be taken into account by introducing two enhancement factors $\kappa_V$ to the seed and $\kappa'_V$ to the sidewalls.
  \end{itemize} 
Altogether, the equation for type-V atoms is written:
\begin{widetext}
\begin{eqnarray} \label{Eq3}
dL_{V}=\left[\frac{2}{R} \frac{1}{U'(\frac{L}{\lambda_{f,V}})}\lambda_{s,V}  \frac{K_{1}(\frac{R}{\lambda_{s,V}})}{K_{0}(\frac{R}{\lambda_{s,V}})} + \frac{2}{R} \frac{U(\frac{L}{\lambda_{f,V}} )}{U'(\frac{L}{\lambda_{f,V}})} \frac{\tan\alpha_{V}}{\pi}\lambda_{f,V} (1+\kappa'_{V})+(1+\kappa_{V})\right] \\ \nonumber
 \times \left[1- \theta_{evap}^{in} \exp(\frac{R_0}{R}) \right] dH_V
\end{eqnarray}
\end{widetext}

The size effect characterized by the parameter $R_0$ describes the enhancement of the evaporation from a small seed (or nanoparticle) and it is often coined as a "Gibbs-Thomson effect". According to Ref.~\cite{Glas2013}, the term "Kelvin effect" is more appropriate, as the Gibbs-Thomson effect impacts actually the nucleation rate at the seed-nanowire interface and can be neglected in our case. Anyway, the characteristic radius is 
\begin{equation} \label{Eq4}
R_0=\frac{2 \gamma \Omega}{k_B T},
\end{equation}
where $\gamma$ is the surface energy of the seed, $\Omega$ the volume of the desorbed atom or molecule (for instance, As$_2$ in the case of arsenides), and $T$ the temperature. The values of $\gamma$ for seeds made of Au, In or Ga, or their alloys, are given in Ref.~\cite{Dubrovskii2009a}, together with the calculated value of the characteristic radius $R_0$ (assuming the evaporation of atoms, not molecules). 

Finally, according to Eq.~\ref{Eq1}, the growth rate of the nanowire during the time interval $dt$ is written
\begin{equation}\label{Eq5}
dL= \min (dL_{III},dL_V).
\end{equation}

\subsection{Dependence on nanowire length}
\begin{figure}[H]
\centering
\includegraphics[page=1,trim=0cm 0cm 0cm 0cm, clip, width=8cm]{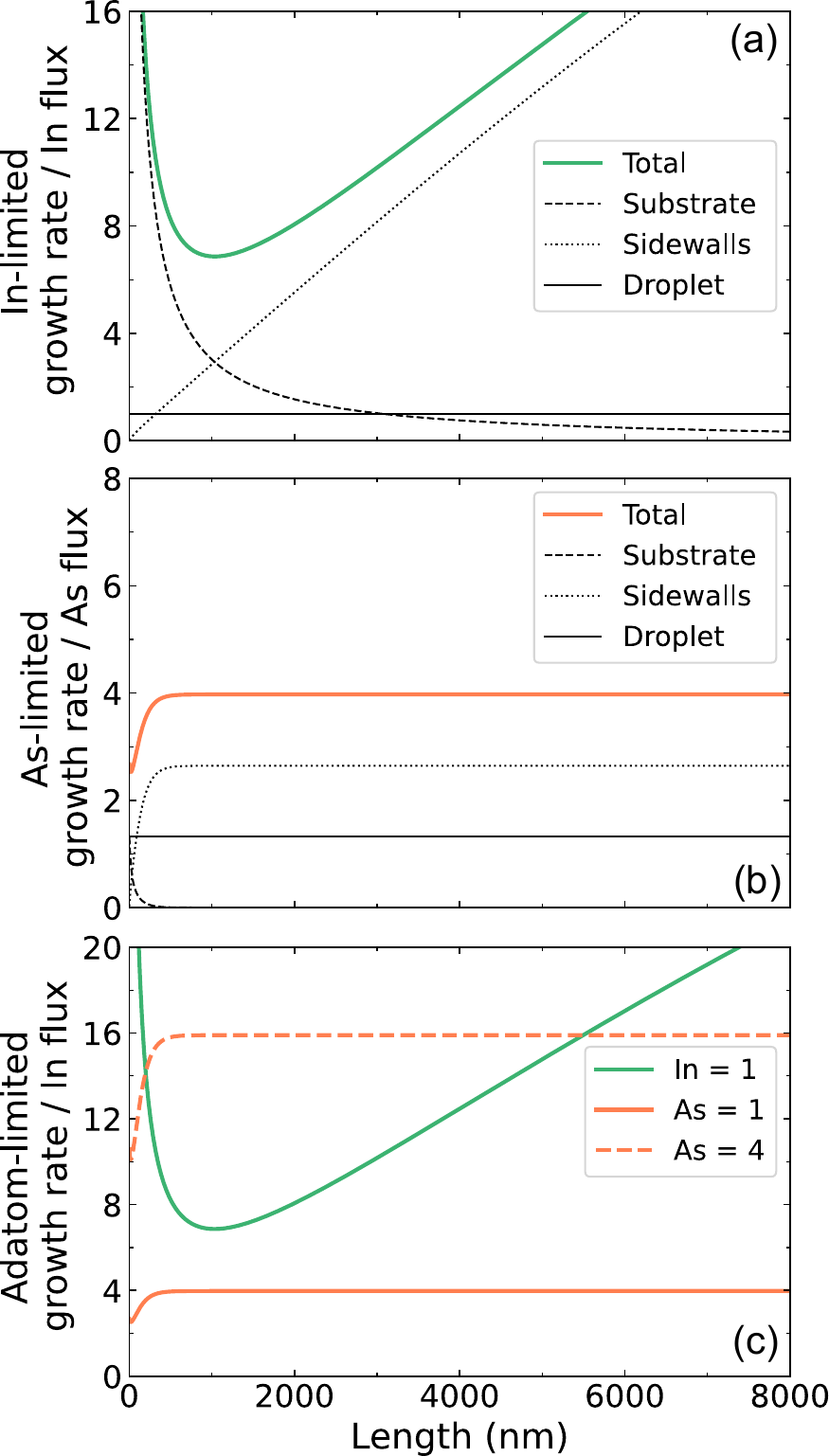}%
\caption{Diffusion limited currents, as a function of the actual nanowire length, for a radius of 25~nm, calculated with the parameters of Table~\ref{table3}. (a) Nanowire growth rate limited by the current of In adatoms, divided by the growth rate of a 2D layer limited by the In flux, $\frac{dL_{In}}{dH_{In}}$ (thick solid green line). The individual contributions of the adatoms diffusing from the substrate (thin solid line), of the adatoms impinging on the sidewalls (dashed line) and of the direct flux to the seed (dotted line) are also shown. (b) Same quantity $\frac{dL_{As}}{dH_{As}}$ for As (thick solid orange line). (c) shows $\frac{dL_{i}}{dH_{In}}$, for $i=$In (green solid line), $i=$As with an As-to-In flux ratio =1 (orange solid line), and $i=$As for an As-to-In flux ratio = 4 (orange dashed line).}
\label{fig:5}
\end{figure}

The diffusion-limited growth rates, Eqs.~\ref{Eq2} and \ref{Eq3}, depend on the radius $R$ and the instantaneous length $L$ of the nanowire. The non-linear character of $L(t)$ has been often commented \cite{Dubrovskii2009b,Glas2010a,Harmand2010}. In the present description, it means that the growth rate of a nanowire of a given radius can be limited by either type-III  or type-V atom species, depending on the growth conditions, but also that the limiting species can change during the growth. This is exemplified in Fig.~\ref{fig:5} for an InAs nanowire with a radius of 25 nm, using the parameters determined in the next section (see Table~\ref{table3}). The In-limited growth rate is plotted versus the nanowire length in green in Fig.~\ref{fig:5}(a), while the As-limited growth rate is plotted in orange in Fig.~\ref{fig:5}(b). For each type of adatom, the different contributions are plotted in grey (substrate, sidewall, seed).

In the case of In, the growth rate decreases very rapidly with increasing nanowire length, reaches a minimum and then increases slowly. This behaviour is explained by the initial contribution of the substrate diffusion which is predominant for nanowire lengths below 1~$\mu$m (due to a short diffusion length on the substrate). Above this value, the current due to the flux onto the sidewalls is predominant and increases even for nanowire lengths beyond $\lambda_{f,In}$. 

In the case of As, the instantaneous growth rate reaches a minimum for very short nanowire lengths (not visible), and increases until reaching rapidly a constant value. Below 200 nm in length, diffusion of As from the substrate and the flux to the nanoparticle are predominant, while above this value, the growth rate due to the seed and the diffusion of atoms impinging the sidewalls (with a very small diffusion length) contribute constantly to the growth rate. 

Under those conditions, one sees that the 25 nm radius nanowire is always limited by As for a V/III flux ratio = 1. For a V/III flux ratio  of 4, this same nanowire sees its growth rate limited alternatively by As and In (first As, then In, then As again), as shown in Fig.~\ref{fig:5}(c). Finally, at sufficiently large V/III flux ratios and typical final lengths of several micrometers, the nanowire elongation rate will be primarily limited by In during growth.

\begin{table}[H]
\caption{Fitting parameters for the study of the As BEP dependence. \label{table3}}
\begin{ruledtabular}
\begin{tabular}{lclc}
In data & &  As data & \\ \hline	 	 	 
$\lambda_{s_{In}}$ (nm) & 200 &  $\lambda_{s_{As}}$ (nm) & 50 \\
$\lambda_{f_{In}}$ (nm) & 7500 &  $\lambda_{f_{As}}$ (nm) & 100\\
$\kappa_{In}$ & 0 &  $\kappa_{As}$ & 6 \\
 & &  $\theta_{evap}^{in}$	& 0.6 \\
 & &  $R_0$ (nm) &	8 \\				
\end{tabular}
\end{ruledtabular}
\end{table}

The limiting aspect of In and As during growth needs to be re-evaluated for each radius. Indeed, the contribution of the sidewalls plays an important role, as the In-limited growth rate curve has a large slope increase with radius reduction. Interestingly, the contribution of the seed to the As-limited growth rate becomes predominant with larger radii (see Fig.~S5 in ~\cite{suppldata2024}).

\subsection{Fitting the experimental data}
We now return to Fig.~\ref{fig:3} which shows the length-radius dependence of four samples grown under the same condition but for the As BEP (see Table I). The sample with the lowest As flux (Fig.~\ref{fig:3}(a)) displays a length dependence on radius which is characteristic of an As-limited growth all along the growth. We use it to determine the values of the As-related parameters. A good fit is obtained (Fig.~\ref{fig:3}(e)) with the parameters of Table~\ref{table3} and the known experimental parameters displayed in Table~\ref{table2}. Similarly, the sample with the highest As flux (Fig.~\ref{fig:3}(d)) shows a rapid decrease of the length when the radius is increased above 20 nm: this is characteristic of a diffusion-limited growth with a long diffusion length, and a good fit is obtained (Fig.~\ref{fig:3}(h)) with the parameters attributed to In in Table \ref{table3}.

By keeping this set of parameters unchanged and adjusting only the As-flux ($H_{As}$, known from thorough calibration performed for each sample), we achieve a good fit for all samples for all values of the nanowire radius. Thus, we confirm the values of the parameters in Table \ref{table3}, and the validity of the approach synthesized in Eq.\ref{Eq5}.

\begin{table}[H]
\caption{Known experimental parameters for the study of the As BEP dependence. \label{table2}}
\begin{ruledtabular}
\begin{tabular}{lcccc}
	%&NW898&	NW899&	NW900&	NW901\\
	&I25R60&	I25R20&	I25R10&	I25R05\\ \hline
In data	 &&&&	 	\\ 	 	 	 
$\alpha_{In}$ ($^\circ$)&	12	&12	&12&	12\\
$H_{In}$  (nm)	&280&	280&	280&	280\\
	&&&&	\\ 					
As data&&&&	\\ 	 	 	 	 	 
$\alpha_{As}$ ($^\circ$)	&38	&38&	38	&38\\
$H_{As}$   (nm)&	3000&	1000&	500&	250\\
\end{tabular}
\end{ruledtabular}
\end{table}

In Fig.~\ref{fig:3}(e)-(h), the dotted lines show the limits of the nanowire elongation versus radius allowed by the As and In currents, respectively, as calculated with our fixed set of parameters  and the calibrated values of the fluxes (Table~\ref{table5}). When the As flux (V/III flux ratio of 1) is low enough, the elongation due to As remains smaller than the elongation due to In for all radii (Fig.~\ref{fig:5}(e)). On the other hand, when the As flux increases to 2, the elongation curve due to As crosses the one due to In. Fig. \ref{fig:3}(f) shows that below 30 nm radius, the elongation allowed by As is systematically smaller than the In one. Above 30 nm radius, the two limits cross and the elongation of the nanowire due to As solely becomes greater with respect to the one due to In. For even higher As fluxes, the critical radius at which the transition between As and In limited regimes shifts toward lower values and the maximum nanowire length increases further.

\begin{table}[H]
\caption{Known experimental parameters for the study of the In BEP dependence. \label{table5}}
\begin{ruledtabular}
\begin{tabular}{lccc}
	%&NW934&NW898&			NW939\\
	&I18R100&I25R60&			I38R42\\ \hline
In data	 &&&	 	\\ 	 	 	 
$\alpha_{In}$ ($^\circ$)&	12	&	12&	12\\
$H_{In}$  (nm)	&		260&280&	330\\
	&&&	\\ 					
As data&&&	\\ 	 	 	 	 	 
$\alpha_{As}$ ($^\circ$)	&38	&38&	38	\\
$H_{As}$   (nm)&	4800&	3000&	3200\\
\end{tabular}
\end{ruledtabular}
\end{table}

Finally, we test our model for different values of the In flux. Fig.~\ref{fig:4}(b) shows the length-radius dependence of three different samples with In fluxes changed by a factor of 1.5 (see Tables~\ref{table4} and~\ref{table5} as well as Fig. S4 in ~\cite{suppldata2024}). We observe that it is possible to keep the same fitting parameters in our model, but for the critical radius $R_0$ (Table~\ref{table6}). An increase in the In flux by a factor of 1.5 leads to the reduction of $R_0$ from 11 nm to 6 nm. In other terms, larger indium fluxes lead to the possibility to grow thinner nanowires.

\begin{table}
\caption{List of $R_0$ for various In fluxes. \label{table6}}
\begin{ruledtabular}
\begin{tabular}{lcccccc}
	%&NW934NW898&		&	NW939\\
	&I18R100&I25R60&			I38R42\\ \hline
$R_0$ (nm)&11&	8&		6\\
\end{tabular}
\end{ruledtabular}
\end{table}

\section{Discussion}

The present study embeds three original aspects that  will be discussed now: 
\begin{enumerate}
    \item we experimentally explore a range of growth conditions (As and In flux) which covers the transition from As-limited to In-limited growth, for a broad range of the nanowire radius; 
    \item we propose a dual-adatom diffusion-limited model which describes the transition, without adjustable parameter, thanks to a systematic preliminary calibration of the In and As flux (see S1 in ~\cite{suppldata2024}); the V/III BEP ratio at the transition strongly depends on the nanowire radius, but also on the instantaneous length; 
    \item the critical radius $R_0$ which characterizes the sharp decrease of the growth rate at small radius is clearly ascribed to the Kelvin effect on As and its value depends on the In flux.
\end{enumerate}

It is interesting to compare the present study to the phase diagram of the growth rate of thin InAs nanowires with respect to the In and As BEP values, Fig.~1 of Ref.~\cite{Babu2011}. Keeping in mind that for the same values of BEPs, the values of the projected flux $F_i \cos \alpha_i$ may differ between two MBE machines, that it depends on the cell positions and, for As, on the nature of the molecular beam, we note that we obtain similar values of the growth rate. Our range of As BEPs covers a similar ratio, but we explore a definitely narrower range of the In flux (a factor of 2). However, by measuring the length-radius dependence for different values of the In and As flux, we obtain a more informative set of experimental data which allows us to discuss quantitatively the mechanisms involved in the growth process.

The dispersion in the length-radius dependence is much bigger than that expected for the Poisson distribution \cite{Dubrovskii2016b,Glas2017,Dubrovski2016}, which may be expected, even under perfectly controlled growth conditions, due to the random nature of step nucleation. The stronger dispersion we observe may be due to an initial delay in the nucleation of nanowires. Such a delay was reported in Ref~\cite{Dayeh2009a} for very short growth times, of the order of the minute, with a dependence on the In flux. Regarding our case, we observe no delay in the apparition of the Bragg peaks in the RHEED pattern, and samples grown for a shorter time (7~min instead of 30~min, not shown) display no nanowires of vanishing length (excepted close to the cut-off) which would have started freshly. The broad range of the nanowire radius associated to each size of the nanocristallite (as mentioned above) suggests that at least the smaller gold nanoparticles move on the substrate (some nanoparticles may even be buried and re-emerge as in GaAs \cite{Dubrovskii2016b,Koivusalo2017}) before growth starts. Hence, we cannot rule out an initial delay, but it is at most of the order of a couple of minutes.

The dispersion may also be due to some roughness of the substrate surface as the growth starts, or to the presence of the 2D layer which progressively grows between the nanowires. Both may influence our determination of the substrate diffusion length $\lambda_s$ which must be considered as an effective capture length. Our value for In ($\lambda_s\approx$ 200~nm) is intermediate between the 10 to 95 nm reported for Ga or In in MBE \cite{Harmand2010,Sibirev2012,Dubrovskii2009a} and the hundreds of nanometers for chemical beam epitaxy \cite{Froberg2007,Jensen2004}. Nonetheless, Fig.~\ref{fig:5}(a), for In, and even more Fig.~\ref{fig:5}(b), for As, show that the substrate contribution is significant only for short nanowires.

Another important contribution is the flux to the seed, either directly from the cell, or due to re-emission from the substrate. The shape of the seed may enhance the contribution of the direct flux, both for In and As. This is the case, for instance, if the shape of the nanoparticle is close to a full sphere. Then the flux collected by the nanoparticle with a large cross-section is used to feed the growth of the nanowire with a much smaller area \cite{Glas2010c}. Although indirectly, post-growth SEM images and TEM images suggest that in the present case, the nanoparticle is approximately half-spherical, independently from the V/III ratio (not shown): then the collection cross-section and the nanowire area are equal. 

As the re-emission of type-III atoms is likely to be negligible (or vanishing \cite{Sibirev2012,Koryakin2019a,Koryakin2019b}), we keep $\kappa_{In}=0$. This can be revisited for substrates on which the metal doesn't stick \cite{Oehler2018, Rieger2012, Krogstrup2010}. The case of As is quite different. The projected flux of As, $F_{As} \cos \alpha_{As}$, is consistently larger than that of In. Consequently, a significant amount of As is not frozen during the growth of the 2D layer. The geometry of the type-V flux has been detailed in Ref.~\cite{Ramdani2013} for As and in more details in Ref~\cite{Pishchagin2021} for P. The effect on the flux to the gold seed may be significant, but the most spectacular aspect from Ref~\cite{Pishchagin2021} is that the diffusion of the type-V atoms along the sidewalls is not as vanishing as it was assumed previously (with the exception of Ref.~\cite{Krogstrup2013}), and the corresponding contribution can be very strongly enhanced by the re-emission from the substrate after adsorption. Our values $\lambda_{f,As}$ and $\kappa_{As}$ (100 nm, and 6, respectively) fit relatively well the experimental data. Nevertheless, $\lambda_{f,As}$ and $\kappa_{As}$ (380 nm, and 0.5, respectively) fit as well the data  (see Fig.~S3 in ~\cite{suppldata2024}) but do not represent the physical behaviour of As (re-emission here is neglected and compensated by a larger (effective) As diffusion length.

We chose a simple description of the size effect, with an overall factor $\left[1-\theta_{evap}^{in} \exp(\frac{R_0}{R})\right]$ in Eq.\ref{Eq3}. A size effect appropriate to each elementary surface of the nanostructure was considered in Ref.~\cite{Dubrovskii2009a} but the experimental data do not allow one to disentangle the different contributions. The characteristic radius $R_0$ for the Kelvin effect in the seed is calculated in Ref.~\cite{Dubrovskii2009a} for mono-atomic evaporation. Using the same parameters, and Eq.~\ref{Eq4} for the evaporation of a di-atomic molecule (As$_2$) \cite{Glas2013}, we obtain $R_0 \approx$~8~nm for a Au particle. Due to a lower surface energy, it is expected to be 2.6 times smaller for pure In, and intermediate for a Au-In alloy. The type-III content in the particle is difficult to measure during growth, yet is assumed to be as high as several 20\%-30\% from post-growth characterization, in strong contrast to a small As content of the order of a few \%, negligible in the calculation of $R_0$. 

The other parameter characterizing the size effect is $\theta_{evap}^{in}$, which is introduced in Ref.~\cite{Dubrovskii2009a} as a ratio of activities of the As in the vapour and in the seed. Its value should be calculated from a complete thermodynamical approach \cite{Glas2010b,Glas2013,Glas2013b,Krogstrup2013}, which is beyond the scope of the present study. Based on Fig.~6 of Ref.~\cite{Glas2013}, which shows that under As-limited conditions the re-evaporated flux is proportional to the incident current to a good approximation, we keep a constant value for $\theta_{evap}^{in}$ under such conditions. This is comforted by the stable value, upon changing the As flux, of the threshold radius below which no nanowire is observed, equal to $\frac{R_0}{\ln(\theta_{evap}^{in})}$.

The last parameter is the diffusion length of In along the nanowire sidewalls. All studies report a long diffusion length, $\lambda_{f,In}$ in the several micrometer range. Our value,  $\lambda_{f,In}=7.5~\mu$m, is based on the non-linear growth rate of long nanowires under As excess (see Fig.~\ref{fig:5}(a)) and provides a lower bound to $\lambda_{f}$. The determination of an upper bound would require to put into evidence the saturation of the growth rate for nanowires definitely larger than $\lambda_{f}$. The observation of cylinder-shaped nanowires is not a relevant criterion: the application of the Burton-Cabrera-Frank model to the propagation of steps along the nanowire sidewalls shows that if the sticking probability of an adatom to a step present on the sidewalls is below a threshold value,  cylinder-shaped nanowires of arbitrary length, larger than the diffusion length limited by desorption, can be obtained \cite{BelletAmalric2020}. The other parameters needed (Table \ref{table2}), are known from the position of the cells and substrate in the chamber, and from measurements of the growth rates of InAs layers (see Fig.~S1 in ~\cite{suppldata2024}).

Having in hand a set of fitting parameters which describes the growth limited by the As minority current, whatever the nanowire radius, and the growth limited by the In minority current, for large values of the radius, we can address the main goal of our study: how do we switch from one limit to the other one, as a function of the two projected fluxes, but also of the nanowire radius and nanowire length, and what is the impact on the growth rate. The calculated limits in Fig.~\ref{fig:5}(a) confirm that for such a V/III flux ratio (=1) the As current is smaller than the In current, whatever the nanowire radius. Fig.~S5 in ~\cite{suppldata2024} confirms that this remains true whatever the nanowire length.

For large nanowire radii, the deep minimum in the In current (which reflects the long diffusion length of In) may fall below the almost constant As current at low V/III flux ratio, as shown in Fig.~S5. Hence the growth starts as As-limited, then switches to In-limited, and back to As-limited for long enough nanowires. 

For a smaller radius, see also Fig.~S5 in ~\cite{suppldata2024}, the As-current is made smaller by the Kelvin effect and the In-limited section shrinks: the growth is As-limited whatever the nanowire length. Thus, without any further assumption, we obtain that for a large enough V/III flux ratio, the growth is indeed In-limited for a large nanowire radius, but remains As-limited for thin nanowires. Thus the length-radius dependence combines the Kelvin effect on As at small radius and the long-diffusion of In at large radius. 

At intermediate radius, the limiting current depends on the nanowire length, so that the transition in the length-radius plot is smooth (see Figs.~\ref{fig:3}(b) and (c)). This remains true when the V/III flux ratio is further increased, as shown in Figs.~\ref{fig:5}(a-b) and Fig.~S5 in ~\cite{suppldata2024}. We note that the minimum radius, given by $\frac{R_0}{\ln(\theta_{evap}^{in})}$, is sharply defined and remains unchanged when changing the As flux. This comforts our assumption that both $R_0$ and $\theta_{evap}^{in}$ do not change under constant In flux and growth temperature.

The evolution of the length-radius dependence, when changing the In flux (Fig.~\ref{fig:4}(b)), is also well described with the same set of parameters, but for $R_0$, see Table ~\ref{table6}. Indeed a decrease of $R_0$ is expected if the In content in the nanoparticle increases, as a result of a smaller surface energy for an In-rich alloy. Unfortunately, although all determinations of the post-growth In content conclude to high values \cite{Harmand2005,Dick2005,Park2006,Zhang2015}, typically 30\% or more, it is quite difficult to assess what was the indium content during the growth. Regarding GaAs nanowires grown by metal-organic chemical vapor deposition, the high value of the Ga content in the Au nanoparticle, and its increase upon increasing the Ga precursor flux, have been confirmed by an \textit{in situ} study of the growth \cite{Maliakkal2019}. A practical result of the shift of the critical radius with In flux is that long, thin nanowires can be grown at high In flux, as already reported for instance in Ref.~\cite{Jung2014}.

A crucial aspect of our approach is that the growth rate, as calculated using Eq.~\ref{Eq5}, depends only on the minority current to the nanoparticle. The currents calculated in Eqs.~\ref{Eq2} and \ref{Eq3} with the fitting parameters determined under a strong excess of the opposite species are a good estimate of the maximum current that one can anticipate for each species. While the minority current defines the nanowire elongation through Eqs.~\ref{Eq5}, no assumption is made for the majority species. Of course, the majority current is expected to ultimately adjust through a specific elimination process. If the minority species is the metal, the As content in the nanoparticle will increase to a value such that the excess As current is eliminated by evaporation. The increase of the As content maintains As-rich conditions, and the increase of $\theta_{evap}^{in}$ above its value in Table~\ref{table3} adjusts the As current contributing to the growth to the same value as the (minority) metal current. The opposite case of As-limited growth is more complex. A symmetric process can take place in gold-seeded growth with an increase of the type-III metal content in the  droplet \cite{Maliakkal2019}. It is probably less efficient than the adjustment of As, and totally ruled out in the case of self-catalyzed growth \cite{Tersoff2015}. An efficient mechanism of back diffusion of Ga along the sidewalls has been identified in the self-catalyzed growth of GaP \cite{Pishchagin2021} and may be present in gold-seeded growth as well. To first order, these two processes do not modify the minority As current and the growth rate. Special care should be taken in more involved case, for instance if the large type-III content in the gold droplet impacts the activity of As, or if a modification in the droplet size occurs during self-catalyzed growth.

The present model describes efficiently our experimental data and helps understanding the mechanisms involved in the growth of compound nanowires. It can be applied to the growth of nanowires made of compound semiconductors, through the VLS, the VSS and the solution-liquid-solid \cite{Laocharoensuk2013} processes, once the equations and parameters describing the dynamics of the two species have been established using strongly unbalanced growth conditions. Indeed, a sharp maximum in the length-radius dependence has been reported for various materials: III-Vs where the type-V atome is likely to be the volatile species, such as GaAs \cite{Dubrovskii2014}, InP \cite{Dubrovskii2014}, InAs \cite{Froberg2007}, or the extreme case of GaN \cite{Maliakkal2016}; but also II-VIs such as CdSe \cite{Laocharoensuk2013}, ZnSe \cite{Laocharoensuk2013} \cite{Rothman2019} and ZnS \cite{Rothman2019} where the cation may be the volatile species like Cd in CdTe \cite{Orru2018}. The role of the Kelvin effect, of nucleation \cite{Dubrovskii2014} and of individual constituents, must be identified thanks to studies under broad growth conditions. 

An obvious extension would be the introduction of equations based on the chemical potentials instead of the adatom density. Note that this should affect primarily the diffusion equation for the majority species (which has to evaporate if not fully consumed by the nanowire growth) while our model uses mainly the diffusion equation for the minority species. However, it would allow us to take into account the nucleation rate \cite{Johansson2019,Dubrovskii2020} and could possibly make the transition in the length-radius plot smoother than in the present model. It should also be possible to detect the effect of the nucleation rate on the growth rate as a function of length measured on a single nanowire with markers \cite{Glas2010a,Harmand2010,Madsen2013}, by choosing a radius at the transition as in Fig.~\ref{fig:5}(c). Another aspect is the decrease of the nucleation probability on a small interface area, which may also play a role in the absence of growth seeded by small Au nanoparticles, although the calculated cut-off appears to be less sharp in the absence of Kelvin effect \cite{Dubrovskii2014,Koryakin2021}. The approach should also be used to describe the nucleation of steps at the nanowire-seed interface, and their propagation \cite{Harmand2018,Maliakkal2020}. Finally, a full thermodynamic assessment of the parameters \cite{Glas2013b,Glas2013,Krogstrup2013} would be clearly a welcome improvement of the present description.

\section{Conclusion}
We developed a generic dual-adatom diffusion-limited model for the VLS and VSS growth of nanowires made of compound semiconductors, that implements the distinct contributions of each species to the nanowire elongation. Knowing the growth parameters of the nanowires in the case each species is in excess, the model considers the growth under intermediate flux ratio by calculating the maximum current to the seed for each species. It allows us to describe the nanowire elongation, when growth is limited alternatively by one of each species - the one which has the smaller current, depending on the radius and the length of the nanowire at the moment of its growth. The model is applied and tested on InAs nanowires, where the growth rate is governed by the smaller current between In and As. It is relevant for other compound semiconductors, self-assembled and gold-seeded nanowires, and should be considered also for understanding the nucleation and propagation of steps at the interface between the seed and the nanowire.

\begin{acknowledgments}
This work was supported by ANR HYBRID (ANR-17-PIRE-0001), ANR IMAGIQUE (ANR-42-PRC-0047), IRP HYNATOQ, the Transatlantic Research Partnership and the LANEF IRGA LABEX BICOQ. We thank Frank Glas for extended discussions on modeling.
\end{acknowledgments}

\bibliographystyle{apsrev4-2}
\bibliography{Dual-adatom-model}

\end{document}

\title{SUPPLEMENTARY DATA FOR\\Dual-adatom diffusion-limited growth model \\for compound nanowires: Application to InAs nanowires}

\author{Danylo Mosiiets}
\author{Yann Genuist}
\author{Joël Cibert}
\affiliation{Univ. Grenoble Alpes, CNRS, Grenoble INP, Institut Néel, 38000 Grenoble, France}
\author{Edith Bellet-Amalric}
\affiliation{Univ. Grenoble Alpes, CEA, Grenoble INP, IRIG, PHELIQS, 38000 Grenoble, France}
\author{Moïra Hocevar}
\affiliation{Univ. Grenoble Alpes, CNRS, Grenoble INP, Institut Néel, 38000 Grenoble, France}

\date{\today}

\maketitle

\section{S1. Indium and Arsenic flux calibration}
We calibrate the flux of In and As atoms emitted by the cells by measuring  the growth rate of InAs layers on an InAs (001) substrate using Reflection High-Energy Electron Diffraction (RHEED) oscillations at different beam equivalent pressures (BEP)  at 420°C. The intensity of a diffraction spot during 2D-layer growth usually features two types of variation: sharp transients due to a change of surface state (stoichiometry and possibly associated reconstructions), for instance when opening or closing a cell shutter, and periodic oscillations reflecting the degree of completion of the surface monolayer, with the period corresponding to the time to complete one monolayer. 

\begin{figure}[H]
\includegraphics[page=1,trim=0cm 0cm 0cm 0cm, clip, width=16cm]{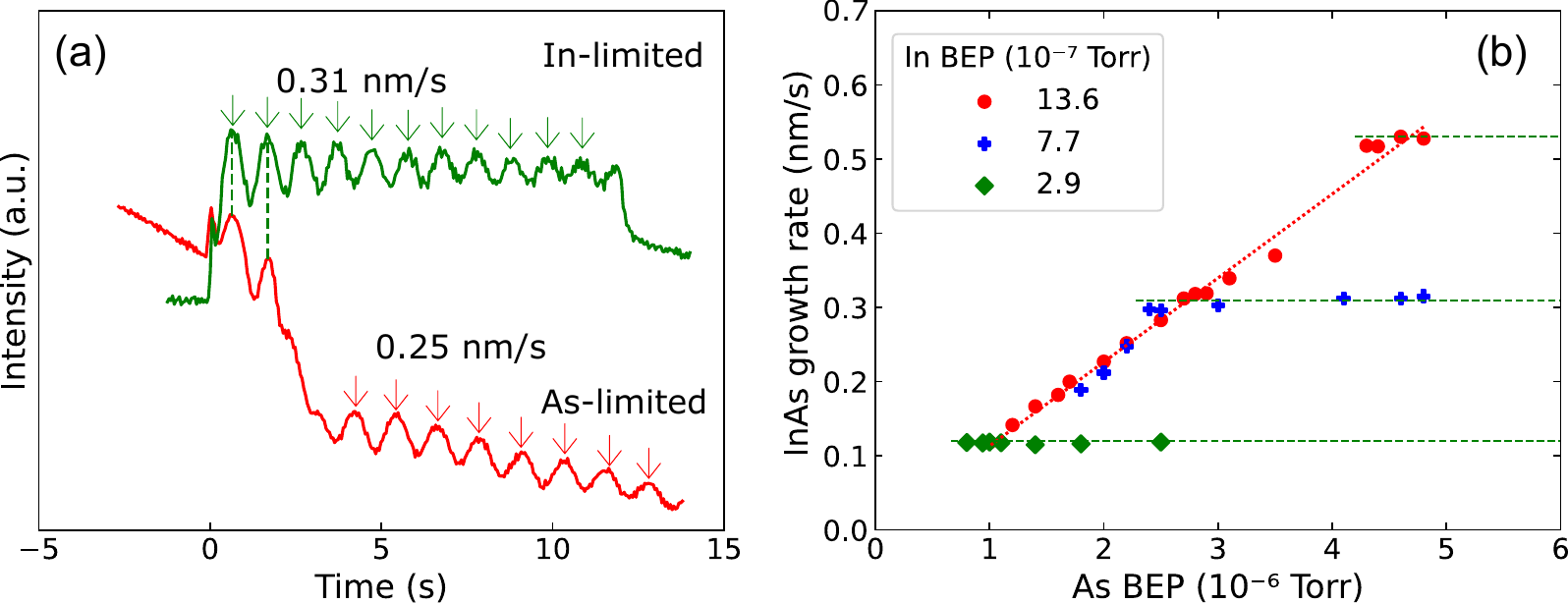}
\caption{(a) RHEED oscillations for InAs at 420$^\circ$C in the Arsenic rich regime (green line, In BEP=7.7$\times \mathrm{10^{-7}}$ Torr and As BEP=4.2$\times \mathrm{10^{-6}}$ Torr) and in the Arsenic poor regime (red line, In BEP=7.7$\times \mathrm{10^{-7}}$ Torr and As BEP=2.2$\times\mathrm{10^{-6}}$ Torr). The arrows show the intensity maxima used to calculate the growth rate in each case. (b) Growth rate versus As BEP for different values of the In BEP: 13.6$\times \mathrm{10^{-7}}$ Torr (red), 7.7$\times \mathrm{10^{-7}}$ Torr (blue) and 2.9$\times\mathrm{10^{-7}}$ Torr (green). Plateaux for In-limited growth rates are shown by dashed lines, and the linear fit at 0.11~nm/s/$\mathrm{10^{-6}}$ Torr for As-limited by the dotted line.}
\label{fig:S1}
\end{figure}

As already well documented in the case of GaAs (Ref [46] in main text), the determination of the growth rate of a III-V layer is quite straightforward in the metal-limited regime, and more subtle in the case of the As-limited regime. The green line in Fig.~\ref{fig:S1}(a) displays the RHEED signal recorded during the growth sequence with an Indium BEP equal to 7.7$\times \mathrm{10^{-7}}$ Torr and a strong As BEP, 4.3$\times \mathrm{10^{-6}}$ Torr. With the As cell kept open, we opened the Indium cell at time=0: In that case, the RHEED intensity featured a sharp transient that we ascribe to a change in the surface state, with a symmetric transient upon closing the In cell. In between, periodic oscillations were observed which we will ascribe below to the In-limited growth rate at Indium BEP 7.7$\times \mathrm{10^{-7}}$ Torr.

It is more delicate to explore the Arsenic limited regime, as growth under Arsenic-poor conditions degrades the InAs surface. Alternatives were described in Ref(Ref [46] in main text) for GaAs. We could nevertheless measure the As-limited growth rate of InAs as follows. We started by measuring the growth rate in the Arsenic rich regime. Then we decreased the As BEP, step by step, keeping constant the In BEP and measuring the RHEED intensity during short sequences with the In cell open. After each sequence, the Indium cell was closed for 2-3 minutes to return to Arsenic-rich conditions and flatten the InAs surface. During this time lapse, Arsenic accumulates on the substrate surface. When the Indium shutter opens again, a complex evolution of the RHEED intensity was observed, as exemplified by the red line in Fig.~\ref{fig:S1}(a): In addition to the sharp transients upon opening and closing the Indium cell, we observe an intensity drift, with different values of the oscillation period before and after the drift. We ascribe the drift to a change of the surface state from As-rich to As-poor as the excess of Arsenic accumulated on the surface is incorporated, giving rise to a rapid oscillation compatible with the In-limited growth rate previously measured. After the drift, the slower oscillations are ascribed to the slower, As-limited growth.

This interpretation is confirmed when plotting the measured growth rates for a full set of measurement at 420°C, Fig.~\ref{fig:S1}(b), for three different values of the In BEP. In the Arsenic rich regime, we see plateaux that correspond to the InAs growth rate limited by Indium. When the As BEP decreases below a threshold, the growth rate is proportional to the As-BEP.

Finally, we extract in Fig.~\ref{fig:S1}(b) the In-limited growth rate (horizontal dashed lines) and the As-mimited growth rate (dotted line with a linear dependence on As BEP). We note that In-limited data points are observed slightly beyond the crossing with the As-limited line: actually, if the values of the As-flux and In-flux, projected onto the surface, are close to each other, the excess As is incorporated slowly, so that the In-limited rate is observed for a long time during the long-lasting transition from As-rich to As-poor.

With a sticking coefficient equal to unity, the flux parallel to the cell axis, in atom nm$^{-1}$s$^{-1}$, is obtained by dividing the measured growth rates by the cosine of the incidence angle, and by the volume of the unit cell in InAs (Ref [46] in main text). It is also possible to calculate the deposited thicknesses, H$_{In}$ and H$_{As}$, by multiplying the growth rate of each adatom with the sample growth time. Note that they do not depend on the orientation.

If the sticking coefficient is smaller, we underestimate the flux. 

Actually, the values of the sticking coefficient has been a matter of debate for GaAs. There is a general agreement that it is close to unity for Gallium. It has been thought for a long time that the As sticking coefficient was limited to $\frac{1}{2}$ for an As$_4$ molecular beam. More recent data suggest that it is larger than $\frac{1}{2}$ under appropriate growth conditions (Ref [63] in main text), and for instance on a (001) surface when clear RHEED oscillations are observed.

In the absence of a fully dedicated study for InAs, we may reasonably assume that the sticking coefficient is close to unity for the conditions of the present calibration (clear RHEED oscillations on (001) surface). Hence we only slightly underestimate the flux values and the $H_i's$. The actual conditions for the growth, with a (111) orientation and formation of nanocrystallites between the nanowires, are certainly not optimized towards limiting the re-emission of As atoms or molecules.

To conclude, we think that the flux values used in the present study are probably slightly, but not dramatically, underestimated, but also that the re-emission of As from the substrate between the nanowires remains high even in the case of the smallest As-BEP.

\newpage

\section{S2. Additional samples from the Arsenic series}

Complement to Figs.~1 and 2 in the main text, showing the two intermediate samples.

\begin{figure}[H]
\includegraphics[page=1,trim=0cm 0cm 0cm 9cm, clip, width=16cm]{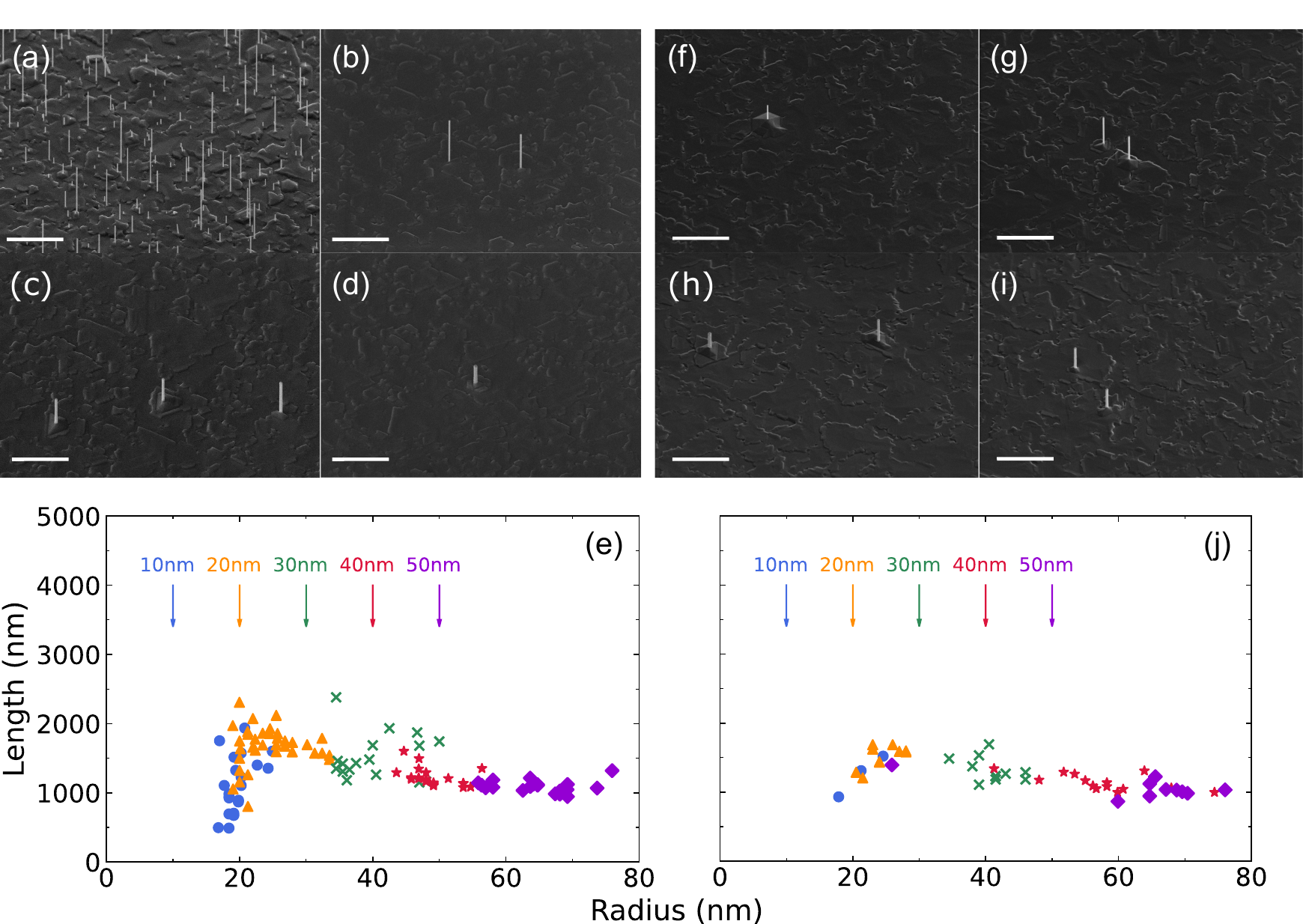}
\caption{Nanowires grown at In BEP 2.5$\times\mathrm{10^{-7}}$ Torr, for As BEPs 5$\times\mathrm{10^{-6}}$ Torr (sample I25R20) and 2.5$\times\mathrm{10^{-6}}$ Torr (sample I25R10). All SEM images are 45$^\circ$ tilted. Scale bars 2 $\mu$m. Radius of colloids: (a) and (f) 10 nm, (b) and (g) 20 nm, (c) and (h) 30 nm, (d) and (i) 40 nm. (e) and (j) Nanowire length $vs.$ radius. The initial radius of colloids is displayed with an arrow, the corresponding nanowires are identified by the color of the symbols.}
\label{fig:S2}
\end{figure}

\newpage

\section{S3. Additional fits of the Arsenic series - test of parameters}

\begin{figure}[H]
\includegraphics[page=1,trim=0cm 0cm 0cm 0cm, clip, width=16.5cm]{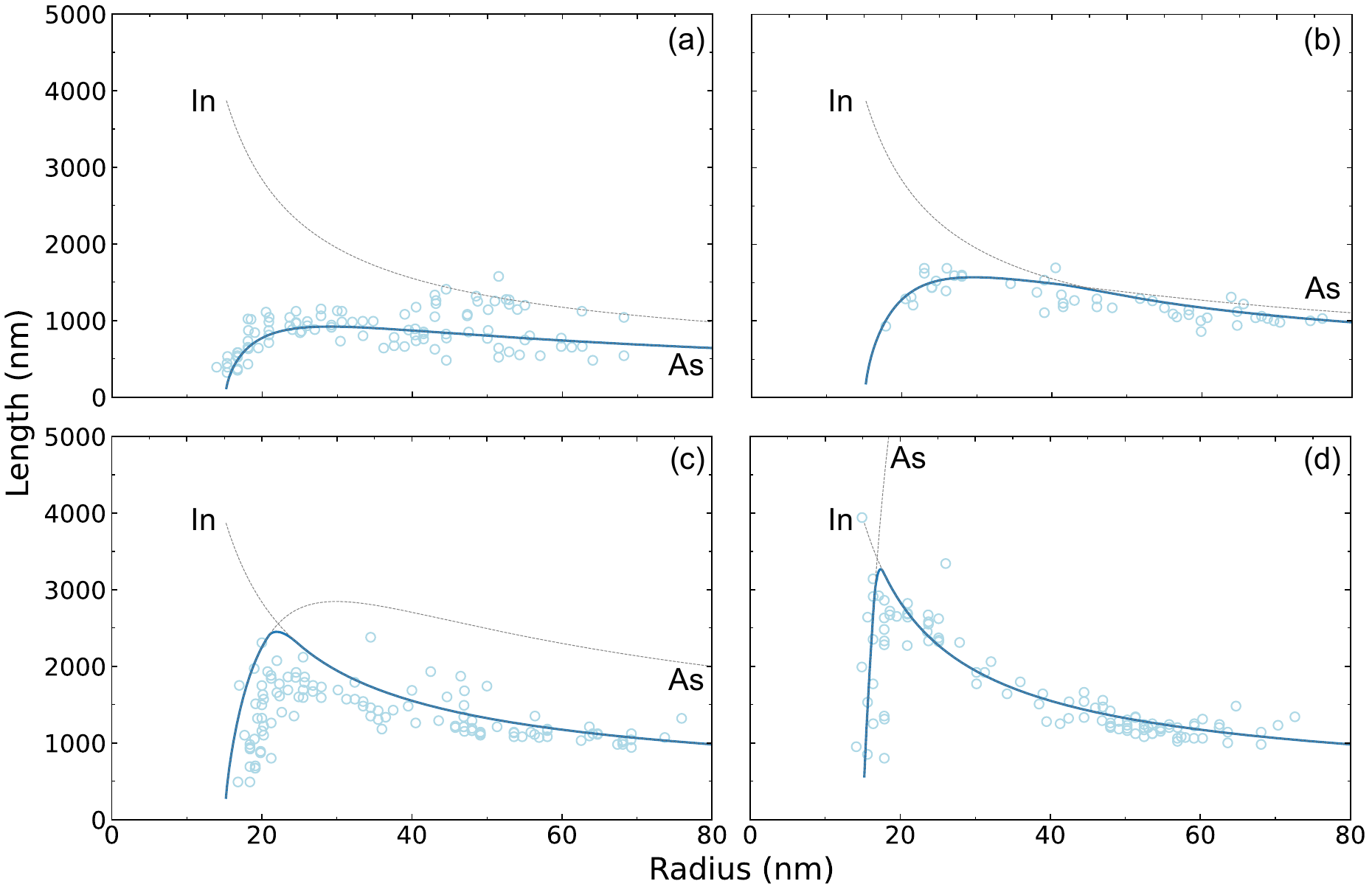}
\caption {Fit of the experimental data for various V/III flux ratios: (a) 5, (b) 10, (c) 20 and (d) 60; The blue full circles show the experimental data, the blue solid lines the fit, the grey dashed lines the length-diameter dependence when growth is solely limited by In and As, respectively. The fitting parameters are $\lambda_{sAs}$ = 280 nm, $\lambda_{fAs}$ = 380 nm, $\kappa_{As}$ = 0.5.}
\label{fig:S3}
\end{figure}

\newpage

\section{S4. Samples from the Indium series}
Complement to Fig.~4 in the main text, showing the details for two samples (the third one, I25R60, belongs also to the As series and is shown in Fig.~1 in the main text).

\begin{figure}[H]
\includegraphics[page=1,trim=0cm 0cm 0cm 0cm, clip, width=16.5cm]{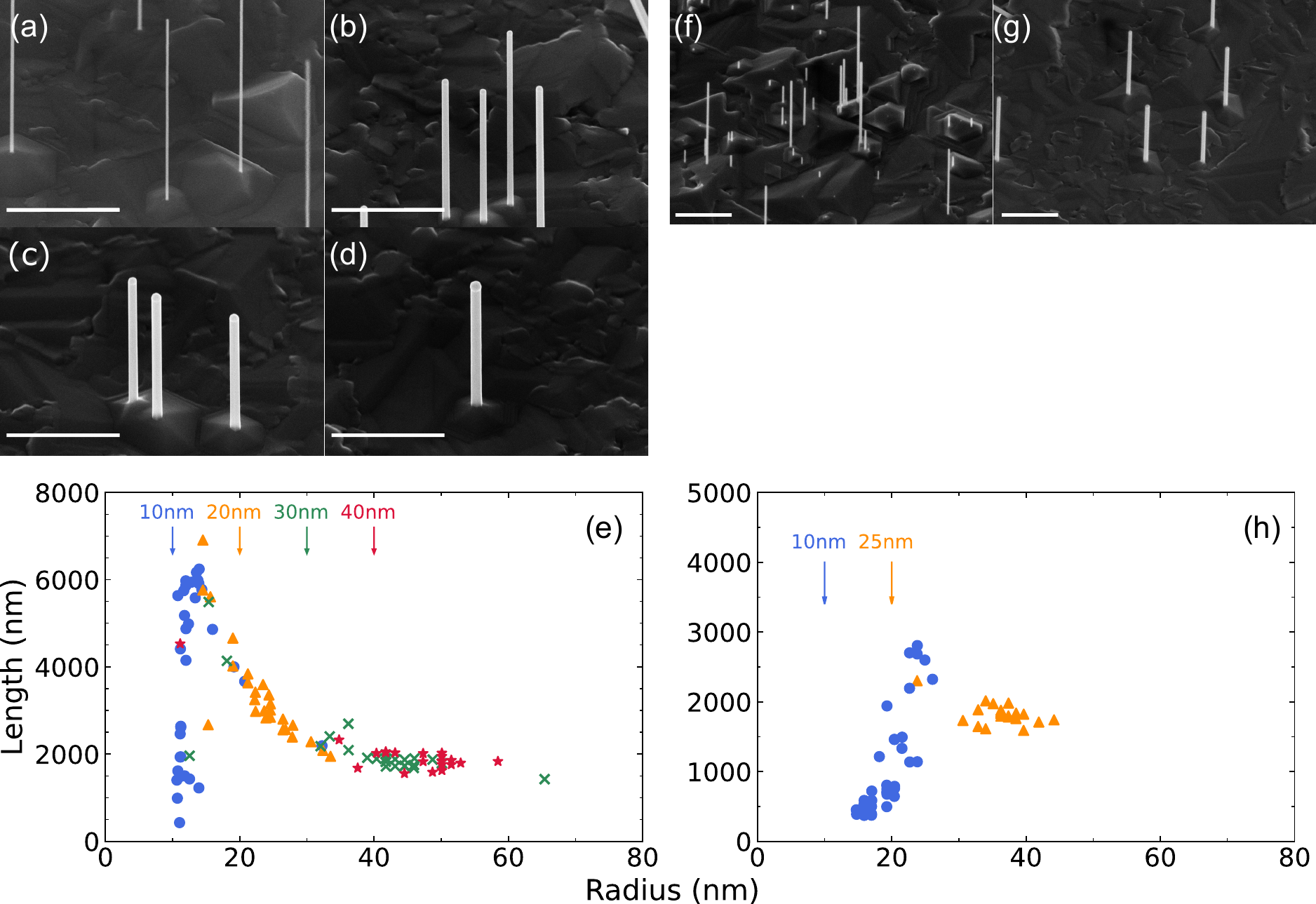}
\caption{Nanowires grown at In BEP: (a-e) 3.8$\times \mathrm{10^{-7}}$ Torr (sample: I38R42), (f-h) 1.8$\times\mathrm{10^{-7}}$ Torr (sample: I18R100). All SEM images with 45$^\circ$ tilt. Scale bars  1 $\mu$m. Radius of colloids: (a) 10 nm, (b) 20 nm, (c) 30 nm, (d) 40 nm, (f) 10 nm and (g) 25 nm. (e) and (h) Length \textit{vs.} radius. The initial radius of colloids is displayed with an arrow, the corresponding nanowires are identified by symbols of the same color.}
\label{fig:S4}
\end{figure}

\newpage

\section{S5. Diameter dependence of the instantaneous growth rate}

Complement to Fig.~5 in the main text, showing the calculated growth rate dependence on nanowire length for different values of the nanowire radius.

\begin{figure}[H]
\includegraphics[page=1,trim=0cm 0cm 0cm 0cm, clip, width=16.5cm]{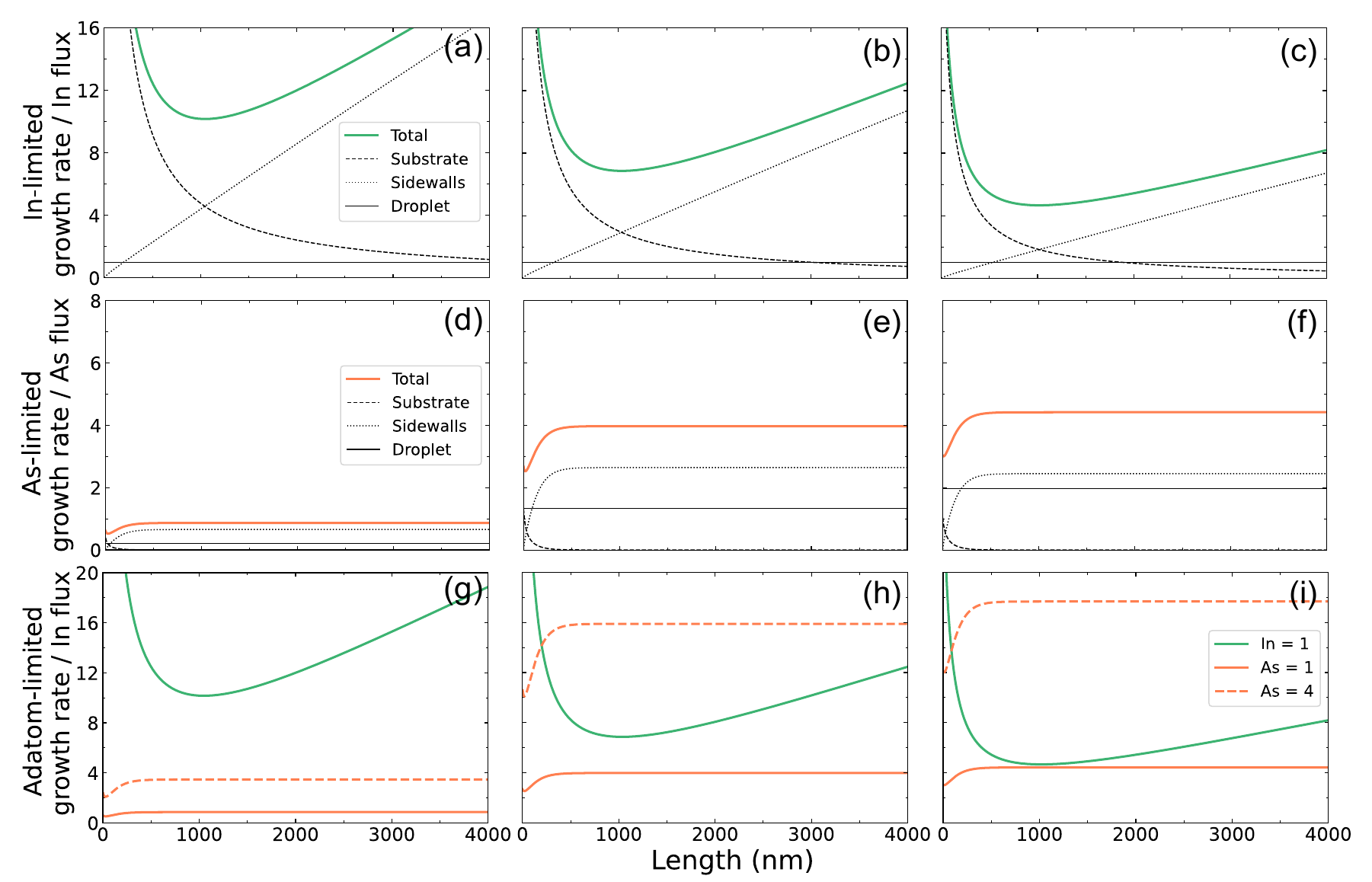}
\caption{Calculated diffusion limited currents for In and As adatoms, as a function of the actual nanowire length, calculated with the parameters of Table I, for a nanowire radius equal to (1st row) 16~nm, (2nd row) 25~nm and (3rd row) 40~nm. In the first line, we plot the growth rate limited by the current of Indium atoms reaching the seed, divided by the growth rate of a 2D layer limited by the Indium flux, $\frac{dL_{In}}{dH_{In}}$ (thick solid green line); The individual contributions of the adatoms diffusing from the substrate (thin solid line), of adatoms impinging on the sidewalls (thin dashed line) and the direct flux to the seed (thin dotted line) are also shown. The second line shows the same quantities for Arsenic, $\frac{dL_{As}}{dH_{As}}$ (thick solid orange line).  The third line shows the In and As-limited currents divided by $H_{In}$, $\frac{dL_{i}}{dH_{In}}$, for $i=$In (green solid line), $i=$As with an As-to-In flux ratio =1 (orange solid line), and $i=$As for an As-to-In flux ratio = 4 (orange dashed line).}
\label{FigS5}
\end{figure}

\newpage

\bibliographystyle{apsrev4-2}
\bibliography{SupplementaryData}